\documentclass{article}[15pt]

\usepackage[dvips]{epsfig}
\usepackage{rotating}

\begin{document}

\title{
Zipf's Law in Importance of Genes for Cancer Classification \\
Using Microarray Data \\
\vspace{0.2in}
\author{
Wentian Li \\
{\small \sl  Lab of Statistical Genetics, Box 192,
Rockefeller University, New York, NY 10021, USA.} \\
}
\date{\today}
}
\maketitle    
\markboth{\sl Li}{\sl Li}

\vspace{-0.2in}

\begin{abstract}
Microarray data consists of mRNA expression levels of thousands of genes 
under certain conditions.  A difference in the expression level of 
a gene at two different conditions/phenotypes, such as cancerous versus 
non-cancerous, one subtype of cancer versus another, before versus 
after a drug treatment, is indicative of the relevance of that gene
to the difference of the high-level phenotype. Each gene can be 
ranked by its ability to distinguish the two conditions. We study 
how the single-gene classification ability decreases with its rank 
(a Zipf's plot). Power-law function in the Zipf's plot is observed
for the four microarray datasets obtained from various cancer studies. 
This power-law behavior in the Zipf's plot is reminiscent 
of similar power-law curves in other natural and social phenomena 
(Zipf's law). However, due to our choice of the measure of importance 
in classification ability, i.e., the  maximized likelihood in a logistic 
regression, the exponent of the power-law function is a function of 
the sample size, instead of a fixed value close to 1 for a typical 
example of Zipf's law. The presence of this power-law behavior is 
important for deciding the number of genes to be used for a 
discriminant microarray data analysis.
\end{abstract}


\section{Introduction}

\indent

Not all of the more than 30,000 human genes and perhaps a multiple 
of that number of proteins [Venter, et al., 2001; IHGSC, 2001] 
are expected to be useful for all situations. For a
specialized phenotype, a particular genetic disease, or a biological 
process of interest, only a portion of the genes are involved. Even 
for the involved genes, their contribution to that phenotype
(or disease, or process) varies: some genes can be more 
important than others. The number of genes involved in a 
phenotype/disease/process also differs from one case to another: 
some phenotypes require a large number of genes to contribute 
(polygenic), others may only need a few (oligogenic). This theme 
has been discussed in the study of human genetic diseases: those caused 
by mutation in one gene are called Mendalian or simple diseases, 
whereas those caused by multiple mutations in many genes, or due to an 
interplay between genetic susceptibility and environmental factors, 
are called complex genetic diseases (see, e.g., [Haines, et al., 1998]).

Once a phenotype, a disease, or a biological process has been
chosen, we may examine each gene's contribution to the
phenotype/disease/process under study, and its ``importance"
can be ranked. This ranking is a very common practice in quantitative
analyses of natural and social phenomena. For example, cities can 
be ranked by their population, companies can be ranked by revenue 
or profit, ecological disasters can be ranked by the population of 
species being destroyed, etc. Plotting the measurement by which 
the ranking is determined, versus the ranking number (1 for the best 
or the largest, etc., 2 for the second-best, $\cdots$) is called a 
Zipf's plot, originated from the work of George Zipf [Zipf, 1935, 1949].

Among the large number of Zipf's plots drawn by Zipf, especially
that of the frequency of word occurrence in human language, it was 
observed that they tend to follow a power-law (algebraic) function: 
$y \sim 1/r^\alpha$, where $r$ is  the rank, $y$ is the measurement 
that determines the rank, and $\alpha \approx 1$ the scaling exponent. 
This power-law behavior is called Zipf's law. Power-law Zipf's plot 
where the value of $\alpha$ is not close to 1 may be called a generalized
form of Zipf's law.  Examples of Zipf's law, besides those mentioned above, 
include magnitude of earthquakes [Sornett, et al., 1996], 
popularity of webpage visits [Crovella \& Bestavros, 1997], 
usage of library books, 
frequency of key word search [Chen \& Leimkuhler, 1986], etc. 
More references on Zipf's law can be found at an online resource at
{\sl http://linkage.rockefeller.edu/wli/zipf}.

This paper addresses the question of whether human genes can
be ranked by their ``importance" in their contribution to a 
particular phenotype such as cancer.  Once a measurement of 
the importance of a gene is defined, we ask whether this measure
of importance as a function of the rank follows an inverse power-law 
function, i.e. Zipf's law.  There are two aspects of this study. 
Qualitatively, Zipf's plot will show whether or not the 
importance of genes, at least for a particular phenotype 
classification task, follows a continuous spectrum. A
discontinuous gap would separate ``important" genes from ``unimportant" 
ones.  Quantitatively, how fast the importance of genes 
falls off as a function of the rank provides information on
how many genes should be kept for this phenotypic classification.

The invention of the microarray (DNA chip) was a breakthrough in
biotechnology (see, e.g., [Southern, 1996; Ekins \& Chu, 1999]), 
which makes it possible 
to monitor the expression level of thousands of gene products such 
as mRNA simultaneously. A comparison of these gene expression levels 
in cancerous as well as normal tissues will point to genes relevant 
to oncogenesis. When each gene is examined individually, the more 
its expression  is different  in cancerous tissues from that in 
normal tissues, the more relevant this gene is to cancer development, 
and the more important it is for cancer classification. In this 
paper, we use microarray data to rank the importance of genes, and 
Zipf's plot of this measure of importance is studied. Section 2
introduces the mathematical formula and quantities used in this paper;
section 3 illustrates the likelihood-rank plots (Zipf's plot) for
the simplest cancer classifications (two-type); section 4 shows
the Zipf's plot for multiple-type classifications; section 5 shows
the Zipf's plot for paired samples classification;  section 6
discusses the issue concerning the exponent in the power-law plot;
and section 7 contains other discussions.

\section{Measure of importance of genes in cancer 
classification: likelihood}

\indent

The measurement we use for the importance of genes in cancer
classification is the maximized likelihood, which is proportional 
to the probability of observing the data when a model is given, 
and when the parameters in the model are adjusted to give the 
maximum value. Mathematically, likelihood $L$ is [Edwards, 1972]:
\begin{equation} 
L = c P(D|M, \theta),
\end{equation} 
where $D$ is all data points in a data set, $M$ is a model, $\theta$ 
represents all parameters in the model, and $c$ is a proportional 
constant (often set to be 1). The maximized likelihood is
\begin{equation} 
\hat{L} = \max_{\theta} L = c P(D|M, \hat{\theta}),
\end{equation} 
where the $\hat{}$ represents a maximization/estimation procedure.
The microarray data sets have the form of 
$\{ x_{1i}, x_{2i}, \cdots, x_{pi}, y_i \}$
($i=1, 2, \cdots N$). Each sample point $i$ (one microarray experiment, 
one tissue sample) contains measurements of (logarithm of ) mRNA expression 
level of $p$ genes ($p$ is typical of the order of thousands), 
and $y$ is a categorical label indicating a condition (e.g. 
cancerous or normal). The raw data in a microarray experiment could be
more complicated: one has to consider background noise, normalization,
and controls.  These considerations depend on the type of chips:
Stanford/cDNA array with two fluorescence dyes [Schena, et al., 1995;
Shalon et al., 1996] or Affymetrix/oligonucleotide arrays with 
only one image intensity to measure but multiple 
oligonucleotide probes [Fodor, et al., 1991]. In this paper, only 
the filtered/processed data are used and the subtle issue of 
scaling/normalization of the raw data is not discussed.

The model $M$ is a classification model (classifier, predictor, 
discriminator, supervised learning machine), that classifies 
the label $y$ by the gene expression level $\{ x_j \}$ (logarithm 
of an image intensity from the DNA chip). We use a particularly 
simple classifier, the single-gene  logistic regression:
\begin{equation}
\label{eq_logit}
 P(y_i=1| x_{ji}) = \frac{1}{ 1+ e^{-a_j - b_j x_{ji}}}.
\hspace{0.1in}
\mbox{$j=1, 2, \cdots p$, and  }
\hspace{0.1in}
\mbox{$i=1, 2, \cdots N.$ }
\end{equation}
In other words, the probability of a sample being in one class is 
a ``sigmoid" or ``logistic" function of the (log) expression level.
If the coefficient $b$ is positive, larger expression levels
lead to higher probabilities of being the $y=1$ label; if
$b < 0$, larger expression level corresponds to the $y=0$ label.

The likelihood of the whole data set is the product of these
model-based probabilities (for a given gene):
\begin{equation}
\label{eq_like_logit}
L_j = \prod_{i=1}^N \left[  P(y=1|x_{ji}) {\bf I}_{y_i=1} +
 (1- P(y=1|x_{ji})) {\bf I}_{ y_i=0}
\right],
\hspace{0.1in}
\mbox{$j=1, 2, \cdots p$ }
\end{equation}
where ${\bf I}$ is the indicator function (1 or 0 depending on
whether the condition is true or not). Eq.(\ref{eq_like_logit}) is 
maximized by adjusting the parameters in the model. The maximized 
likelihood $\hat{L}$ can then be used to rank all genes: the larger
the $\hat{L}$, the higher the ranking.

\section{Zipf's plot of classification likelihood}

\indent

The binomial (two-class) logistic regression Eq.(\ref{eq_logit})
has been applied to two microarray data sets. The first is colon
cancer data from Princeton University [Alon, et al., 1999]. The
expression levels of 2000 genes were available for 62 tissue samples:
40 cancerous and 22 normal. The second data set is leukemia
data from Whitehead Institute/MIT [Golub, et al., 1999]. Expression
levels of 7129 genes were measured on 72 leukemia samples: 47 obtained 
from tissues of one subtype of leukemia,  acute lymphoblastic 
leukemia (ALL), and 25 from tissues of another subtype of leukemia,
acute myeloid leukemia (AML). Although the number of labels is 2 for 
both datasets, we distinguish cancerous from normal tissues in the 
first data set, whereas distinguish one cancer subtype from another
in the second data set.

For the colon cancer data, the single-gene maximized likelihood
Eq.(\ref{eq_like_logit}) for each gene was calculated and ranked.
The likelihood-rank plot (Zipf's plot) is shown in Fig.1 in log-log 
scale. The power-law behavior of the curve is clearly visible.
Fitting the first 600 genes using a generalized form of Zipf's law:
\begin{equation}
\label{eq_zipf}
\hat{L}_r \sim 1/r^\alpha,
\end{equation}
leads to exponent $\alpha \approx 2.18$, whereas the fitting of
the top 1000 genes leads to $\alpha \approx 2.10$. The genes from
roughly rank 1000 to 2000 do not seem to follow the same
power-law decay of the likelihood. As will be discussed more
later in this paper, the exponent $\alpha$ is not an intrinsic
quantity for our likelihood-rank plots. The reason is that the 
likelihood is a product of probabilities of $N$ sample points 
$L_r  \propto (p_1 p_2 \cdots p_N)_r \sim \bar{p}_r^N$; if the 
per-sample averaged likelihood $\bar{p}_r$ of rank-r gene does 
not change with the sample size $N$, the exponent $\alpha$ is 
then a function of $N$: $\alpha \sim  -N \log(\bar{p}_r)/log(r)$.

Fig.2 shows the Zipf's plot of the second data set. Among the 72 
samples, 38 are from bone marrow tissues, and were separated as a
training set in [Golub, et al., 1999]. These 38 samples were
considered to be more homogeneous, while the rest of the samples
were from various sources or other tissue types such as peripheral
blood [Golub, et al., 1999], and may not be homogeneous. From Fig.2,
the Zipf's plot obtained from the training set seems to follow 
a generalized form of Zipf's law with a fitting exponent 
of $\alpha \approx 2.56$ from the top 900 genes. The Zipf's plot
of all sample points seems to deviate from a power-law trend
around rank 100-200. As mentioned earlier, the exponent $\alpha$ 
from a bigger data set is indeed larger than the exponent from 
a small data set.

When the top genes are examined, it was found that the top-ranking
genes for the training set [Li \& Yang, 2001] and those for the 
overall set [Li, et al., 2001] may not be identical. For example,
the top performing gene for the 38-sample training set is No.4847,
{\sl zyxin}, a gene encoding the LIM domain protein used
in cell adhesion in fibroblasts [Golub, et al., 1999]. On the other
hand, the best performing gene for the 72-sample combined set is
No. 1834, {\sl CD33 antigene} which encodes cell surface proteins
commonly found in AML leukemia cells (see, e.g., [Lauria, et al., 1994]). 
The zyxin becomes the rank-4 gene for the combined dataset.  This 
difference reflects a certain degree of inhomogeneity between samples 
in the training set and those in the remaining set (testing or 
validation set). Note from Fig.1  that for the training set, 
genes from rank 3 to 9 exhibits similar likelihood, and form a flat 
step on the Zipf's plot. Such steps are a dominant feature in 
the Zipf's plot of the frequency of word occurrence in randomly 
generated texts [Li, 1992].

\section{Classification of multiple cancer classes}

\indent

The logistic regression Eq.(\ref{eq_logit}) for a two-class data set
can be generalized to multiple classes: multinomial logistic
regression [Agresti, 1996]:
\begin{equation}
P(y_i=I|x_{ji}) = \frac{e^{ -a_{I} -b_{I} x_{ji} }}
{\sum_{K=1}^C e^{-a_{K} - b_{K} x_{ji}}}
\hspace{0.2in}
\mbox{$j=1, 2, \cdots p$,  $I=1,2, \cdots C$, $i=1,2, \cdots N$,}
\end{equation}
where the label $y$ can be in one of the $C$ classes, and there are
two parameters for each class (though only $C-1$ class probabilities
are independent). A gene is important (higher maximized likelihood)
if it is more able to distinguish all $C$ classes simultaneously.

For this multiple-class analysis, we use the microarray data for
lymphoma (Stanford University and National Cancer Institute
[Alizadeh, et al., 2000]).  There are a total 96 tissue samples, with
66 cancerous and 30 normal. Within the cancer samples, there are
46 diffuse large B-cell lymphoma (DLBCL), 9 follicular lymphoma (FL), 
11 chronic lymphocyte leukemia (CLL). These 3 cancer subtypes
plus the normal are the 4 classes to be distinguished. 
In [Alizadeh, et al., 2000], it is also recommended that two new
subclasses of DLBCL can be defined based on the microarray data.
These are the germinal centre B-like DLBCL (GC-DLBCL) and the
activated B-like DLBCL (A-DLBCL). The GC-DLBCL, A-DLBCL subclasses 
plus the normal can be the 3 classes to be distinguished.

Fig.3 shows the Zipf's plot of multinomial logistic regression
likelihood for both the 4-class and the 3-class classification.
The Zipf's plot for the 3-class multinomial logistic regression
follows a perfect power-law function for {\sl all} genes.
No deviation from power-law behavior was observed even  for 
the low-ranking genes. The Zipf's plot for the 4-class
multinomial logistic regression, on the other hand, does
not seem to follow a power-law function. The fall off near
the rank-200 gene is perhaps due to the computer roundoff error
since the value of likelihood at rank-200 is already as low
as $10^{-39}$ (the likelihood is multipled by $10^{10}$ in Fig.3). 
The fitting of the top 200 genes by a power-law function is, 
however, very good.

\section{Cancer treatment effect}

\indent

Another interesting situation is provided by the data set from 
Stanford University and the Norwegian Radium Hospital [Perou, 
et al., 2000].  Part of the data is the expression level of 8102 
genes measured before and after a 16-week course of doxorubicin 
chemotherapy [Perou, et al., 2000] on 20 patients. Naively, one 
may consider these 40 microarray experiments as an example 
of binary logistic regression. However, logistic regression 
in Eq.(\ref{eq_logit}) does not apply because it requires 
samples to be independent of each other, whereas microarray experiments 
done on the same patient are clearly not independent.  This situation 
can be handled by the logistic regression of paired case-controls 
(case is a sample with label 1, and control is a sample with label 0) 
[Breslow \& Day, 1980]:
\begin{equation}
\label{eq_pair}
P(x(case)_{ji} - x(control)_{ji}) = 
\frac{1}{1 + e^{ b_j (x(case)_{ji}- x(control)_{ji} )}},
\hspace{0.1in}
\mbox{$j=1, 2, \cdots p$, $i=1,2, \cdots N$. }
\end{equation}
Note that the probability is no longer that for observing $y$ -- when 
a case sample and a control sample are paired, their $y$ value is 
fixed at 1 and 0 -- but that of observing the difference of $x$'s. 
Also note that the first parameter $a$ is now zero.

Fig.4 shows the Zipf's plot for genes in the breast cancer microarray
data. The top three genes were able to perfectly identify the chemotherapy 
effect -- the expression level is always higher (or lower) 
before the treatment than after in all 20 samples. The likelihood
of a perfect fit is equal to 1. Unlike the previous three plots,
the likelihood of genes in Fig.4 does not follow a power-law function,
or even a smooth function.  There seems to be a gap from the ranking 
of 19 to the ranking of 20-25.  Fitting the two segments (from 4 to 19, 
and from 25 to 500) by power-law functions leads to different exponents 
(2.51 versus 1.77).

Plots like Fig.4 can be good news for a microarray data analysis,
because there seems to be a separation between ``relevant" and 
``irrelevant" genes. Irrelevant genes are not important for 
distinguishing samples before and after chemotherapy, and may 
be discarded for further analysis. Of course, this is only a 
rough description of the gene set. A more systematic approach can 
be based on the framework of model selection (see, e.g., [Burnham \& 
Anderson, 1998]) or model averaging (see, e.g., [Geisser, 1993]). 
With model selection, the expression level of different genes can be 
combined, and adding one gene increases the number of parameters in 
the model by 1. The appropriate number of genes to be included in 
a classification is the model with the best ``model selection criterion"
such as Akaike information criterion [Burnham \& Anderson, 1998] 
or Bayesian information criterion [Raftery, 1995], with a best balance 
between a larger likelihood value and fewer numbers of parameters 
[Li \& Yang, 2001, Li, et al., 2001]. With model averaging, there 
is in principle no limitation on the number of genes to be included, 
but irrelevant genes have smaller weights in an averaged classifier
[Golub, et al., 1999, Li \& Yang, 2001]. This makes the effective 
number of genes used much smaller than the apparent number.

\section{Scaling exponent}

\indent

As mentioned earlier, the exponent of the inverse power-law fitting 
function for Zipf's plot (Figs. 1-4) depends on the sample size. 
The reason for this is that the measurement for the importance
of a gene, the sample classification likelihood, is a function of 
the sample size. Here we ask the question whether one can define 
a normalized exponent.  Since $\hat{L}_r \sim 1/r^\alpha$, 
we can obviously draw Zipf's plot for the per-sample likelihood: 
$\bar{p}_r = \hat{L}_r^{1/N} \sim 1/r^{(\alpha/N)}$. There is another 
consideration for classification of more than two labels: it is 
unfair to compare per-sample classification likelihoods when the 
number of classes differs.  Just by random guess, the probability of 
classifying the label correctly for a binary-class case is 0.5, 
whereas that for classifying $C$ labels is $1/C$. We can normalize 
the per-sample likelihood $\bar{p}_r$ by the random-guess probability:
$\hat{l}_r = \bar{p}_r/(1/C) \sim C/r^{(\alpha/N)}$ with the same 
scaling exponent.

Fig.5 shows the Zipf's plot of normalized per-sample likelihood
$\hat{l}_r$ for all data sets analyzed so far. These curves can be 
compared in two ways. First, the performance of the top-ranking genes
can be compared by their $\hat{l}_r$'s.  It is clear from Fig.5 that 
for the leukemia data, the performance obtained from the training set
is better than that obtained from the whole set, indicating a possible
heterogeneity in the data set. The performance of the top genes 
for the breast cancer data set is better (in terms of classification)
than that for the leukemia data set, which in turn is better than 
that for the colon cancer data set. 

It is usually difficult to compare performance between a two-label 
classification and a multiple-label classification, because it 
depends on the base-line expectation. Two base-line expectations 
(called null models) were defined in [Li \& Yang, 2001, Li, et al., 2001]: 
one is to randomly guess all classes with equal probability, and 
another is to guess the class by the proportion of samples with 
this class in the data set. In Fig.5, the first base-line expectation
is used as the normalization factor. The performance of the 4-label 
classification relative to this base-line expectation for the 
lymphoma data is considered to be better than that of the 3-label 
classification, partly due to its low expectation. 

Different data sets in Fig.5 can also be compared by the rate of
falling of likelihood. The fall-off rate, 
as measured by the scaling exponent $\alpha$, ranges from 3 to 9 
per 100 samples (i.e., $\alpha/N \approx 0.03 - 0.09$).  It 
is interesting that  scaling exponents obtained from the leukemia 
data (both the whole data set and the training set) and from the 
lymphoma data (two DLBCL subtypes plus normal) are almost identical
(around 65 - 67 per 100 samples). The breast cancer data is different 
from other data sets for falling faster in the likelihood-rank plot.

\section{Discussion}

\indent

Results from Figs. 1-4 established that when microarray data is 
used to classify cancer tissues, the classification likelihood of 
individual genes follows a generalized form of Zipf's law. The 
reason that these power-law functions in the Zipf's plot do not 
have a scaling exponent equal to 1 is partly because exponent 
$\alpha$ in Eq.(\ref{eq_zipf}) depends on the sample size $N$. 
Besides the issue of the exponent value, the overall power-law 
trend is excellent: a perfect power-law function for all points in 
Fig.3 (3-label classification, in 3.5 decades), a partial power-law 
fitting in a range of 2.5 decades (Fig.1), 3 decades (Fig.2), and 
2 decades (Fig.3, 4-label classification), respectively. The 
only poor fitting by a power-law function is Fig.4. The Zipf's plot 
is flattern out for low-ranking genes in Figs. 1 and 2, while 
it drops off in Fig.3 (4-label classification).  Which functional 
form is more generic for low-ranking genes is not clear, 
perhaps our results are sample-size sensitive.

Our Zipf's plot can be compared to those obtained from biomolecular
sequences. In [Gamow \& Y\u{c}as, 1955], a Zipf's plot for 20 
amino acids usage in protein sequences was presented (in linear-linear 
scale). Redrawing their data in log-log scale does not show any 
power-law behavior (result not shown). Recently, it was claimed 
that oligonucleotide frequencies in DNA sequences follow Zipf's
law [Mantegna, et al., 1994]. This paper drew criticism on its strong 
claim concerning the connection between Zipf's law and human language 
[Martindale \& Konopka, 1995; Israeloff, et al., 1996, Bonhoeffer, et al., 
1996a,1996b, Voss, 1996]: one of the best counter-examples is Zipf's law 
in money-typing texts [Li, 1992]. Also, the paper did not show 
convincingly that the Zipf's plot for oligonucleotide usage was better 
fitted by a power-law function [Martindale \& Konopka, 1996]: 
the deviation from the power-law fitting function can be gradual 
and systematic, an indication that the power-law function is not 
the best choice of fitting function. Finally, the scaling exponent 
in the power-law fitting function in [Mantegna, et al., 1994] is much 
smaller than 1 [Li, 1996]. In comparison, our Zipf's plots in Figs 1-5 
are a much better example of a generalized form of Zipf's law than those 
in [Gamow \& Y\u{c}as, 1955] and [Mantegna, et al., 1994].

One may wonder whether the power-law behavior in Figs 1-5 can be 
derived by a simple random model. In [Gamow \& Y\u{c}as, 1955], the 
Zipf's plot of the frequency of amino acids usage was compared to a 
``random partition of a unit length" model. Suppose a unit interval 
is randomly partitioned into $p$ segments (e.g. $p=20$ for 20 amino 
acids). These segments are ranked by their size $L_{(1)}= max_i(L_i), 
L_{(2)}= max_i (L_i \in L - L_{(1)}) \cdot L_{(p)}= min_i(L_i)$, etc. 
If this random partition is repeated, the mean value of the ranked 
size can be shown to be [Gamow \& Y\u{c}as, 1955]:
\begin{equation}
\label{eq_gamow}
<L_{(r)}> = \frac{1}{p} \sum_{i=1}^{p+1-r} \frac{1}{p+1-i} 
 \hspace{0.1in} \mbox{ $r=1,2, \cdots p$}
\end{equation}
Drawing $<L_{(r)}>$ from Eq.(8) versus rank $r$ shows that it is a 
straight line in linear(y)-log(x) plot, and not a straight line in
either log-log or log(y)-linear(x) plots (results not shown). Actually
this analytic result may not be applicable to our case, because
we keep track of the performance of a given gene on all samples,
whereas in Eq.(\ref{eq_gamow}) the longest interval (or any given
ranking interval) is averaged over all random simulation. In
any case, the power-law behavior in Fig.1-5 does not seem to
be explainable by a simple random model.

We may ask whether Zipf's plot has any practical application for 
microarray data analysis. In information retrieval [van Rijsbergen, 
1975, Salton, 1988] and library/documentation science [Egghe \& 
Rousseau, 1990], Zipf's law is an important foundation that many 
applications are based upon. It is one of the ``bibliometric laws" 
[White \& McCain, 1989] concerning regularities in bibliographies, 
lists of authors, citation lists, etc. For the purpose of finding relevant, 
content-bearing  words (``keywords"), common (highest-ranking) and 
rare (lowest-ranking) words should be avoided [Luhn, 1957,1958].

Do we have a similar situation where the highest-ranking genes
may not be interesting for cancer classification?  (Lowest-ranking 
genes are obviously not interesting.) Our ranking system is not
really the same as  for word usage, since a discrimination
or classification ability has already been included in our 
definition, whereas it is not included in the word usage example. 
In this sense, our top ranking genes are in fact most efficient
for the purpose of cancer classification. On the other hand,
if one is more interested in subtle gene effects, not in the
known main/dominant effect, it is perhaps useful to remove the 
well-known genes from the list and examine other genes in a future
study.  This idea was not tried in our previous analysis [Li \& Yang, 2001,
Li, et al., 2001].

In conclusion, a generalized form of Zipf's law was observed in
microarray data for the likelihood on cancer classification using
single-gene logistic regression. We suspect that this power-law
behavior is generic rather than an exception. A rank-likelihood plot 
(Zipf's plot) can be a useful quantitative tool for discriminant 
microarray data analysis.

\vspace{0.1in}
\noindent
{\bf Acknowledgements:} The author would like to thank Yaning Yang 
for discussion on the random partition of unit interval, Ronald Rousseau 
and Heting Chu for suggesting references on library sciences, Victoria 
Haghighi and Joanne Edington for help with the microarray data.
The work was supported by NIH grants K01HG00024 and HG00008.


\section*{References}

\vspace{0.07in}
\noindent
Agresti A (1996),
{\sl An Introduction to Categorical Data Analysis} (Wiley \& Sons).

\vspace{0.07in}
\noindent
Alizadeh AA, Eisen MB,  et al.  (2000),
``Distinct types of diffuse large B-cell lymphoma identified by
gene expression profiling",
{\sl Nature}, 403:503-511.

\vspace{0.07in}
\noindent
Alon U, Barkai N, Notterman DA, Gish K, Ybarra S, Mack D,
Levine AJ (1999), ``Broad patterns of gene expression revealed by
clustering analysis of tumor and normal colon tissues probed by
oligonucleotide arrays",
{\sl Proceedings of National Academy of Sciences}, 96(12):6745-6750.

\vspace{0.07in}
\noindent
Bonhoeffer S, Herz AVM, Boerlijst MC, Nee S, Nowak MA, May RM (1996a)
``Explaining 'linguistic features' of noncoding DNA",
{\sl Science},  271(5245):14-15.

\vspace{0.07in}
\noindent
Bonhoeffer S, Herz AVM, Boerlijst MC, Nee S, Nowak MA, May RM (1996b),
``No signs of hidden language in noncoding DNA" (letters),
{\sl Physical Review Letters}, 76(11):1977.

\vspace{0.07in}
\noindent
Breslow NE, Day NE (1980),
{\sl Statistical Methods in Cancer Research. I - The Analysis
of Case-Control Studies} (International Agency for
Research on Cancer, Lyon).

\vspace{0.07in}
\noindent
Burnham KP, Anderson DR (1998),
{\sl Model Selection and Inference} (Springer).

\vspace{0.07in}
\noindent
Chatzidimitriou-Dreismann CA, Streffer RMF, Larhammar D (1996),
``Lack of biological significance in the 'linguistic features' 
of noncoding DNA-a quantitative analysis",
{\sl Nucleic Acids Research}, 24(9):1676-1681.

\vspace{0.07in}
\noindent
Chen YS, Leimkuhler FF (1987), 
``Analysis of Zipf's law: an index approach", 
Information Processing and Management, 23:71-182. 

\vspace{0.07in}
\noindent
Crovella ME, Bestavros A (1997), 
``Self-similarity in world wide web traffic: evidence and 
possible causes", IEEE/ACM Transactions on Networking, 
5(6):835-846. 

\vspace{0.07in}
\noindent
Edwards AWF (1972), {\sl Likelihood} (Cambridge Univ Press).

\vspace{0.07in}
\noindent
Egghe L, Rousseau R (1990),
{\sl Introduction to Informetrics: Quantitative Methods in
Library, Documentation and Information Science} (Elsevier). 

\vspace{0.07in}
\noindent
Ekins R, Chu FW (1999),
``Microarrays: their origin and applications", 
{\sl Trends in Biotechmology}, 17:217-218.

\vspace{0.07in}
\noindent
Fodor SP, Read JL, Pirrung MC, Stryer L, Lu AT, Solas D (1991), 
``Light-directed, spatially addressable parallel
chemical synthesisLight-directed, spatially addressable parallel
chemical synthesis",
{\sl Science}, 251:767-773.

\vspace{0.07in}
\noindent
Gamow G, Y\u{c}as M (1955),
``Statistical correlation of protein and ribonucleic acid composition",
{\sl Proceedings of National Academy of Sciences}, 41(12):1011-1019.

\vspace{0.07in}
\noindent
Geisser S (1993), {\sl Predictive Inference: An Introduction}
(Chapman \& Hall).

\vspace{0.07in}
\noindent
Golub TR, Sonim DK, Tamayo P, Huard C, Gassenbeek M,
Mesirov JP, Coller H, Loh ML,  Downing JR, Caligiuri MA,
Bloomfield CD, Lander ES (1999),
``Molecular classification of cancer: class discovery
and class prediction by gene expression monitoring",
{\sl Science}, 286:531-536.

\vspace{0.07in}
\noindent
IHGSC (International Human Genome Sequencing Consortium) (2001),
``Initial sequencing and analysis of the human genome",
{\sl Nature}, 409:860-921.

\vspace{0.07in}
\noindent
Israeloff NE, Kagalenko M, Chan K (1996),
``Can Zipf distinguish language from noise in noncoding DNA?"
(letters) {\sl Physical Review Letters}, 76(11):1976.

\vspace{0.07in}
\noindent
Haines JL, Pericak-Vance MA (1998), eds.
{\sl Approaches to Gene Mapping in Complex Human Diseases}
(Wiley-Liss).

\vspace{0.07in}
\noindent
Lauria F, Raspadori D, et al. (1994),
``Increased expression of myeloid antigen markers in adult acute 
lymphoblastic leukemia patients: diagnostic and prognostic implications",
British Journal of Haematology, 87:286-292.

\vspace{0.07in}
\noindent
Li W (1992),
``Random texts exhibit Zipf's-law-like word frequency distribution",
{\sl IEEE Transactions on Information Theory}, 38:1842-1845.

\vspace{0.07in}
\noindent
Li W (1996),
``Comments on 'Bell curves and monkey languages' " (letters to the editor), 
{\sl Complexity}, 1(6):6. 

\vspace{0.07in}
\noindent
Li W, Yang Y (2001),
``How many genes are needed for a discriminant microarray data 
analysis?", Proceedings of the Critical Assessment of Microarray         
Data Analysis Workshop (CAMDA 2000).

\vspace{0.07in}
\noindent
Li W, Yang Y, Edington J, Haghighi F (2001), 
``Determining the number of genes needed for cancer classification
using microarray data", submitted.

\vspace{0.07in}
\noindent
Luhn HP (1957),
``A statistical approach to mechanized encoding and search of
literature information",
IBM Journal of Research and Development, 1:309-317.

\vspace{0.07in}
\noindent
Luhn HP (1958),
``The automatic creation of literature abstract",
IBM Journal of Research and Development, 2:159-165.

\vspace{0.07in}
\noindent
Mantegna RN, Buldyrev SV, Goldberger AL, Havlin S,
Peng CK, Simon M, Stanley HE (1994),
``Linguistic features of noncoding DNA sequences",
{\sl Physical Review Letters}, 73:3169-3172.

\vspace{0.07in}
\noindent
Martindale C, Konopka AK (1995),
``Noncoding DNA, Zipf's law, and language" (letters),
{\sl Science}, 268(5212):789.

\vspace{0.07in}
\noindent
Martindale C, Konopka AK (1996),
``Oligonucleotide frequencies in DNA follow a Yule distribution",
{\sl Computers and Chemistry}, 20:35-38.

\vspace{0.07in}
\noindent
Perou CM, Sorlie T, et al. (2000),
``Molecular portraits of human breast tumors",
{\sl Nature}, 406:747-752.

\vspace{0.07in}
\noindent
Raftery AE (1995), ``Bayesian model selection in social research",
in {\sl Sociological Methodology}, ed. PV Marsden (Blackwells),
pages 185-195.

\vspace{0.07in}
\noindent
Salton G (1988),
{\sl Automatic Text Processing: The Transformation,
Analysis, and Retrieval of Information} (Addison-Wesley).

\vspace{0.07in}
\noindent
Schena M, Shalon D, Davis RW, Brown PO (1995),
``Quantitative monitoring of gene expression patterns with a 
complementary DNA microarray", {\sl Science}, 270:467-470.

\vspace{0.07in}
\noindent
Shalon D, Smith SJ, Brown PO (1996),
``A DNA microarray system for analyzing complex DNA samples
using two-color fluorescent probe hybridization", 
{\sl Genome Research}, 6:639-645.

\vspace{0.07in}
\noindent
Sornette D, Knopoff L, Kagan YY, Vanneste C (1996), 
``Rank-ordering statistics of extreme events: application to 
the distribution of large earthquakes", Journal of 
Geophysical Research, 101(B6):13883-13894.

\vspace{0.07in}
\noindent
Southern EM (1996),
``DNA chips: analysing sequence by hybridization to oligonucleotides 
on a large scale", 
{\sl Trends in Genetics}, 12:110-115.

\vspace{0.07in}
\noindent
van Rijsbergen CJ (1975),
{\sl Information Retrieval} (Butterworths).

\vspace{0.07in}
\noindent
Venter JC (2001), {\sl et al.} ``The sequence of the human genome",
{\sl Science}, 291:1304-1351.

\vspace{0.07in}
\noindent
Voss RF (1996),
``Linguistic features of noncoding DNA sequences - Comment"
(letters), {\sl Physical Review Letters}, 76(11):1978.

\vspace{0.07in}
\noindent
White HD, McCain KW (1989),
``Bibliometrics",
{\sl Annual Review of Information Science and Technology}, 24:119-186.

\vspace{0.07in}
\noindent
Zipf GF (1935), {\sl Psycho-Biology of Languages}
(Houghton-Mifflin)

\vspace{0.07in}
\noindent
Zipf GF (1949), {\sl Human Behavior and the Principle of Least Effort}
(Addison-Wesley).

\newpage
\begin{figure}
\begin{center}
  \begin{turn}{-90}
  \epsfig{file=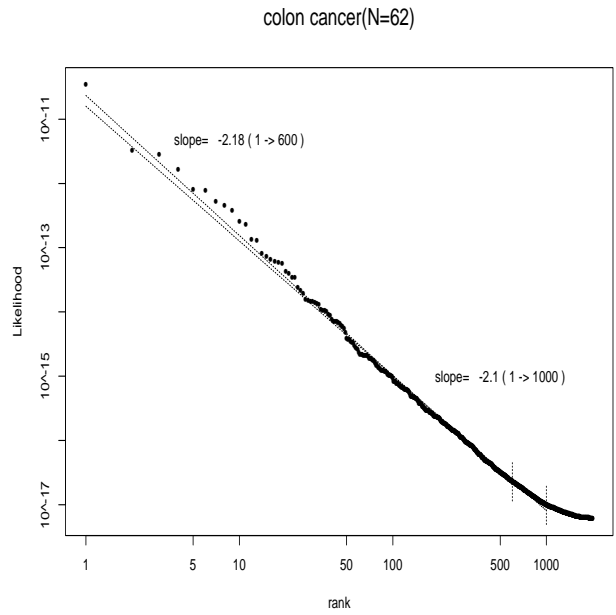, height=8cm, width=8cm}
  \end{turn}
\end{center}
\caption{
Zipf's plot for colon cancer and normal tissue classification: 
maximized likelihood of single-variable logistic regression (Eq.(3)) 
for the top performing genes vs their ranks (in log-log scale).
}
\label{fig1}
\end{figure}

\begin{figure}
\begin{center}
  \begin{turn}{-90}
  \epsfig{file=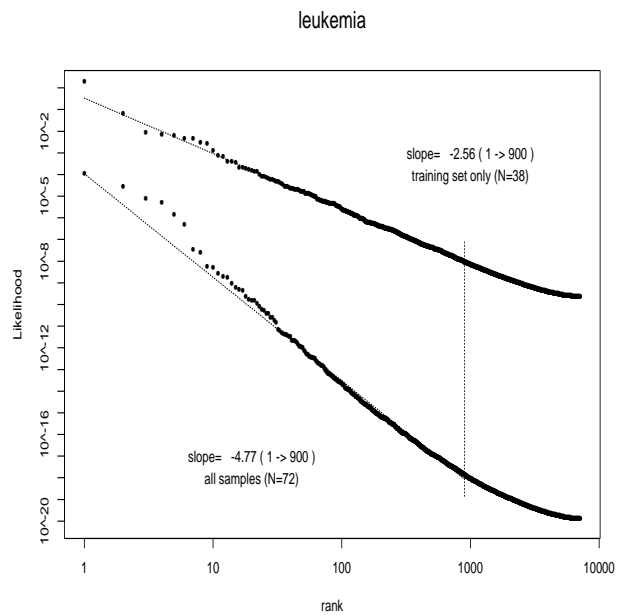, height=8cm, width=8cm}
  \end{turn}
\end{center}
\caption{
Zipf's plot for leukemia subtype classification:
maximized likelihood of single-variable logistic regression
(Eq.(3)) for the top genes vs their ranks, in log-log scale.
The upper line is obtained from the ``training set" which
contains 38 samples, and the lower line is from the 
``training plus testing set" which contains 72 samples.
}
\label{fig2}
\end{figure}

\newpage
\begin{figure}
\begin{center}
  \begin{turn}{-90}
  \epsfig{file=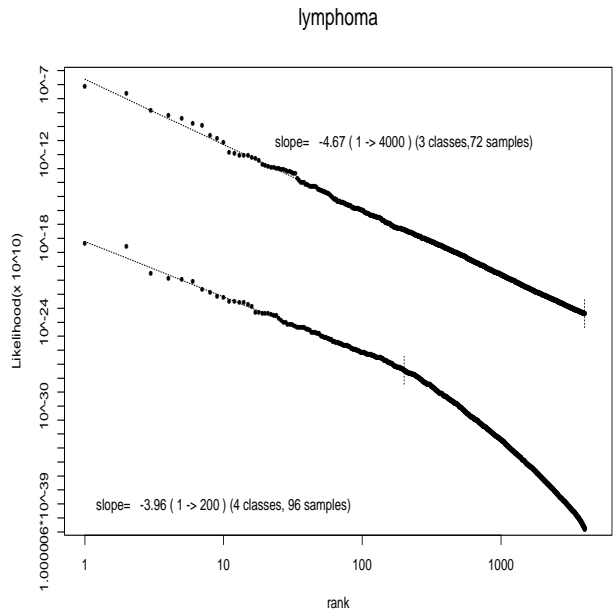, height=8cm, width=8cm}
  \end{turn}
\end{center}
\caption{
Zipf's plot for lymphoma subtypes classification:
maximized likelihood of single-variable multinomial logistic 
regression (Eq.(6)) for top-ranking genes vs. their ranks,
in log-log scale. The upper line is for the 3-class classification
(GC-DLBCL, A-DLBCL, normal), and the lower line is for the
4-class situation (DLBCL, FL, CLL, normal).
}
\label{fig3}
\end{figure}

\begin{figure}
\begin{center}
  \begin{turn}{-90}
  \epsfig{file=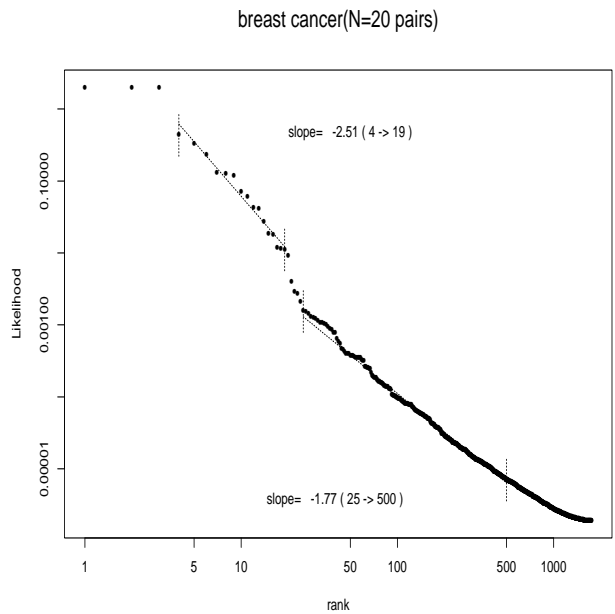, height=8cm, width=8cm}
  \end{turn}
\end{center}
\caption{
Zipf's plot for the breast cancer treatment effect:
maximized likelihood of single-variable paired case-control
logistic regression (Eq.(7)) for top genes vs. their rank
(in log-log scale).
}
\label{fig4}
\end{figure}

\newpage
\begin{figure}
\begin{center}
  \begin{turn}{-90}
  \epsfig{file=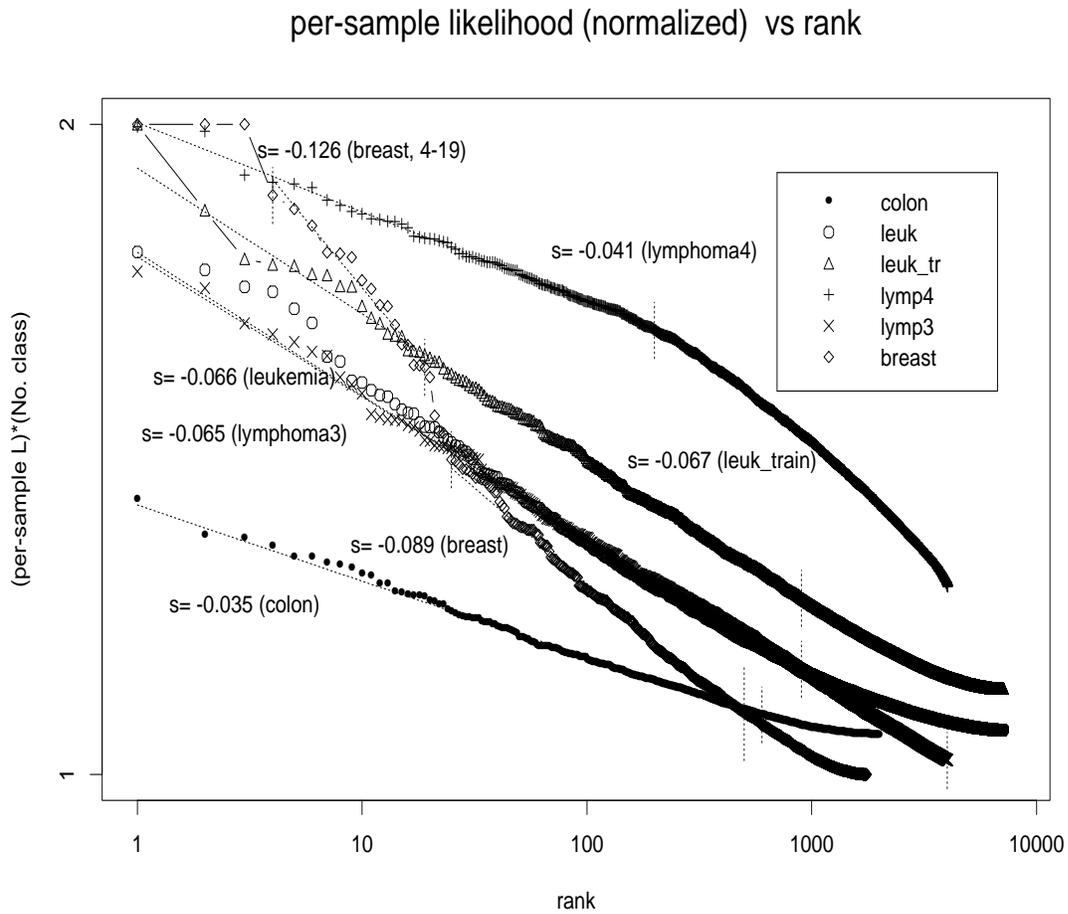, height=14cm, width=12cm}
  \end{turn}
\end{center}
\caption{
Normalized, per-sample maximum likelihood ($\hat{l}_r = \hat{L}_r^{1/N}/ (1/C)$)
where $C$ is the number of classes (e.g., $C=2$ for binomial logistic
regression) for top genes vs. gene ranks, for all data sets. The
corresponding $\hat{L}_r$ vs rank plots are in Figs. 1-4.
}
\label{fig5}
\end{figure}

\end{document}